# Compressing the Data Densely by New Geflochtener to Accelerate Web

Hemant Kumar Saini
Research Scholar
Department of CSE
RTU, Kota, India

Satpal Singh Kushwaha
Associate Professor
Department of CSE
MITRC, Alwar, India

C. Rama Krishna, Ph.D
Associate Professor
Department of CSE
NITTTR, Chandigarh, India

## ABSTRACT
At the present scenario of the internet, there exist many optimization techniques to improve the Web speed but almost expensive in terms of bandwidth. So after a long investigation on different techniques to compress the data without any loss, a new algorithm is proposed based on LZ77 family which selectively models the references with backward movement and encodes the longest matches through greedy parsing with the shortest path technique to compresses the data with high density. This idea seems to be useful since the single Web Page contains many repetitive words which create havoc in consuming space, so let it removes such unnecessary redundancies with 70% efficiency and compress the pages with 23.75 - 35% compression ratio. This also helps in reducing the latencies over Web while transmitting the large data of MB's in seconds over the 10 MBps connection. The proposed method has also been compared with other gzip compatible compressors on the three different compression corpora such as Calgary, Canterbury and enwik8 that proves the success of the work.

## General Terms
Data Compression, Geflochtener Algorithm, Iterative Compression, Lempel-Ziv Variants.

## Keywords
Shortest path technique, iterative compression longest matching, greedy parsing, backward references, Web compression, HTTP

## 1. INTRODUCTION
Compression is the diminution of the physical size of information block to save space and transmission time. Compression can be done on just the data or the entire packet including the header. Data compression is a technique of elimination of all extra spaces, inserting a single repeater to indicate the repeated bytes or characters and replace smaller bits for frequent characters.

Compression is of two types: lossless compression and lossy compression [1] [2]. Lossless compression reforms a compressed file similar to its original form. On the other, hand lossy compression removes the unnecessary data but can't be reproduced exactly. There exist many old and new algorithms for lossless compression which are to be studied e.g., LZ77 [3], LZSS [3], Zopfli [4].

## 2. SOME EXISTING METHODS FOR LOSSLESS DATA COMPRESSION
In this section, some of the existing LZ77 variants and compression algorithms are studied and analyzed their limitations.

### 2.1 LZ77 Sliding Window Algorithm
LZ77 compresses the data by replacing the repeated occurrences with the reference to single copy in the uncompressed input stream. A match is found in lookaheadBuffer which is to be encoded by the length-distance pairs. To stop such matches, the compressor keeps the track of most recent data in a structure called a window. With this window starts to slide at the end and proceeds backwards as the compression is predominated and the window will terminate its sliding, if a sufficient length is matched or it may correlate better with next input.

While (! empty lookaheadBuffer)

{
    get a remission (position, length) to longer match from search buffer;
    if (length>0)
    {
    Output (position, length, nextsymbol);
    transpose the window length+1 position along;
    }
    else
    {
    Output (0, 0, first symbol in lookaheadBuffer);
    transpose the window 1 position along;
    }
}

Search Buffer is the storage which has recently matched encoded characters [3]. LookAheadBuffer is also the storage which contains remaining part of characters that would be matched with SearchBuffer [5]. Tuple is the combination of ($o$, $l$, $c$) where $o$ represent the offset i.e. the bytes from LookAheadBuffer which match in Search buffer, $l$ is the length of the match and $c$ is the next byte to be matched. But the problem occurs, if the sequences of character repeated are larger than the size of search buffer this will decline its performance as the text goes out of the entry even found in the LookAheadBuffer and it won't be considered in matching.

### 2.2 LZSS
This is an alternative of LZ77 which is based on the dictionary encoding technique. It replaces a string of symbols with the remission to a dictionary position for the similar string. In comparing to LZ77 where the dictionary references may be longer than the search buffer, LZSS omits such references. Furthermore, LZSS adds a one-bit flag which represents whether the succeeding lump of data is a literal (byte) or a referral to an offset/length pair.

```
While (! empty lookaheadBuffer)
{
        get a pointer (position, match) to the longest match;
        if (length > MINIMUM_MATCH_LENGTH)
        {
                output (POINTER_FLAG, position, length);
                transpose the window length characters along;
        }
        else
        {
                output (SYMBOL_FLAG, first symbol of lookaheadBuffer);
                transpose the window 1 character along;
        }
}
```

LZSS removes the inclusion of next non-matching byte into each word. This algorithm needs offset and length as references. It also includes an extra flag at each step to find the output tuple which indicates matched length or a single symbol. It yields a better performance over the LZ77 Compression algorithm.

## 2.3 Iterative Compression

After analyzing many compression techniques, the main difficult task in all those would be the choosing of good set of representative rows. Recurring with one iteration to the next, new representative rows may be selected, and old ones are to be discarded [6]. Even though the representative rows may keep changing, each iteration monotonically improves the global quality. In fact, for many cases, even a small number of iterations may be sufficient to deliver significant compression performance. Furthermore, each iteration of the algorithm requires only a single scan over the data, leading to a fast compression scheme.

**Input:** A table T, a user specified value k and an error tolerance vector e.

**Output:** A compressed table $T_c$ and a set of representative rows $P = \{P_1,\ldots,P_k\}$

 Pick a random set of representative rows P

 While totalcov (P, T) is increasing do

  {

   For each row R in T, find $P_{max}$ (R)

   Recomputed each $P_i$ in P as follow:

   {

    For each attribute $X_J$

    $P_i[X_J] = f\ v\ (X_J, G(P_i))$

   }

  }

In this each row R in T is assigned to a representative row $P_{max}(R)$ that gives the most coverage to among the members of P. On the next step a new set of representative rows is computed. Here the sliding window of size $2*e_j$ is then moved along these sorted micro-intervals to find the range that is most frequently matched. In this all discussion it is found that by varying the representative rows the compression ratio betters but there is still a limitation of variation because with the increment of rows there is a situation where it could reduces the CPU cycles. Hence, after experimenting variations it is considered that it should be limited to 100 for the best as it should not either increase the time slices as well as nor decreasing the compression.

## 2.4 Limitation

As seen in LZ77, all the characters are encoded into length and match even the non-matched characters. Search buffer is taken much longer than LookAheadBuffer. And the non-matched pairs waste the space by encoding them as length and offset. So the new LZSS was published to modify the LZ77 which encodes only a pointer when the string is longer than the pointer itself. Hence it sends the bit before each symbol to find whether it is a pointer to a character. Again with all the above considerations Google gives a new heuristic by finding all the possible backward configurations including the non-backward references also and takes the shortest one. But it still uses the previous length in its output and appends every time the length and distance to the LZ77 arrays even when the length is less than the shortest matched string. This consumes the space by again fetching the array for the process. So the modifications are proposed that it will output the tuple like in original LZ77 to save the space and use recent length for better compression, which will be described in the next section.

## 3. EMPIRICAL APPROACH

HTTP compression is the capability of the Apache server for the better bandwidth and greater transmission speeds over the web [7]. Any data which is sent from the server in compressed form is called as HTTP data. And the browsers that support compression use mostly two schemes: gzip and deflate [8]. Both the schemes try to encode the content using LZ77 algorithm. Next in the new researches of Google, adds a new lossless Zopfli compression technique developed by Jyrki Alakuijala and Lode Vandevenne [9, 10].

Unlike the LZ-77 and Zopfli, the proposed system does not concern about the emptiness in the dictionary when the window slides over the data until it fills lookaheadBuffer. Hence use < *0, 0, store* > to encode the characters in store that does not match in dictionary [11].

On the other hand, lengthscore is introduced in the output tuple which is a couple of length and distance. This leads to search the best sequence from the longest match for the better efficiency compresses small characters before the large ones. This compresses the data with high density saving the space on web server.

.

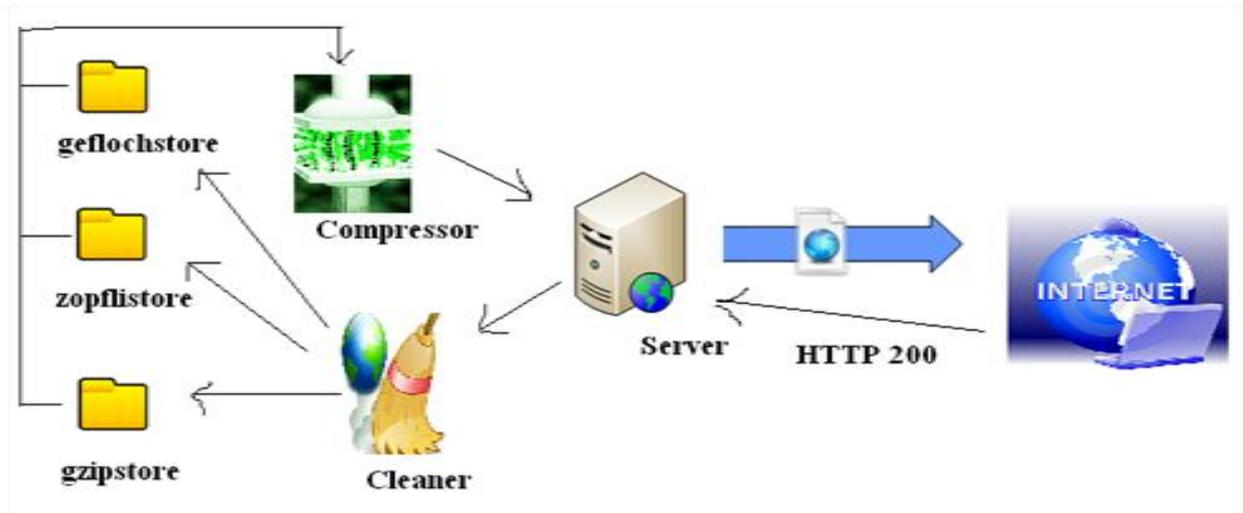

**Fig 1: Scenario of the Proposed System**

## 4. PROPOSED SYSTEM
The proposed system is quite effective than the earlier mentioned techniques. The method is based on iterative entropy based model and a shortest path search to find a low bit cost through all possible representations. Hence to overcome such issues over length in Zopfli as mentioned in section 2.4, it is proposed to compare the lengthscore itself in place of the previous lengths in tuple. Also, for the large unmatched characters it considers only the single context of LZ77Store to append the length and distance to an array.

The system consists of a compressor whose algorithm is discussed in section 4.1 and a Cleaner to flush off the compressed streams. As and when the user searches for his/her query, the server tries to search in pages and respond with those, but the proposed system is being implemented in between this process only without any extra setup. Whenever the fetched pages are ready to send they first pass through the proposed system where the data being compressed in a format compatible to the browser and then send over the Internet. This like when the Internet responds with the HTTP "O.K" message for complete transfer, the system decides to clean the storage of the compressed files with the cleaner so that they do not waste the space by having the copies of original data. The complete scenario is illustrated in Fig.1.

### 4.1 Proposed Geflochtener Algorithm
for (i = instart; i < inend; i++)
{
    **Maximum amount of blocks to split into 100**;
    Update the sliding hash value;
    Find the longest matched cache;
    Gets a score of the length given the distance;
    if (lengthscore >= MIN_MATCH)
      {
        Verifies if length and dist are indeed valid, assert;
    **output StoreLitLenDist(lengthscore, dist, store);**
        shift the window length characters along;
      }
    else
    {
      **output StoreLitLenDist(0, 0, store);**
      shift the window 1 character along;
    }
}

In this algorithm, instart is the starting position of the window, inend is the end position of the window size, Litlen contains the literal symbols or length values i.e. literal per length, lengthscore is the length itself, dist indicates the distance and MIN_MATCH is the shortest distance in length. Literal symbols and the distance both are about the same size.

### 4.2 Compression Process
The page content (leaving images) passes through a litlendistance generator where page data converted into literals with the details of their length and distance. Next these literals and lengths pass through scorer where each literal get the score on the basis of the distance. Now these would be verified before the matching of the literals. As these are validated then they get transfer to the iterator where the bytes are compared to get the longest match from the backward references. As and when it gets longest match they put into a Longest Matched Cache (LMC) and decides on the shortest paths with the best length first. Then as the best lengths are firstly scanned the greedy heuristic verifies the lengthscore and clears the length now shifts the window slider for the next matches. Likewise, the matched pairs after traversing all the paths transmits the matched phrases to the entropy encoder which resembles the Huffman tree bits to value of symbols and get down to compressed stream which will be transmitted over Internet with the header bits set to content encoding gzip.

### 4.3 Compression Strength
For detailed analysis, let us assume the input stream as "ABCDEFGHIJ" at very start so it has not found any backward references. When the window slides the next stream strikes "ABCDEFGHIJ" then it evaporates the distance and length as -10 where '-'shows the back movement. Likewise when it strikes with "AAAAAFGHIJ" as shown in Table 1, it has multiple choice for the references to encode A's with distance and length as – (1) -1 and 4 (2) -10 and 5 respectively. Next when "AAADEFGHIJ" comes into track then to encode 'AAA' it gets back matched with distance -8 and length 3; distance -9 and length 3 and distance -10 and length 3 each with different possible tuple. Likewise, the largest of distances are being considered which is more probable statistically and leads to smaller entropy.

**Table 1. Selection of backward references**

| Line | Input Stream | Distance (d) | Length (l) |
|---|---|---|---|
| 1 | ABCDEFGHIJ | not available | not available |
| 2 | ABCDEFGHIJ | -10 | 10 |
| 3 | AAAAAFGHIJ | -1 | 4 |
|   | AAAAAFGHIJ | -10 | 5 |
| 4 | AAADEFGHIJ | -8 | 3 |
|   |  | -9 | 3 |
|   |  | -10 | 3 |

## 4.4 Algorithm Analysis

With the proposed Geflochtener algorithm, it is tried to compress the data by encoding phrases from lookahead buffer as references in sliding window so that the lookahead buffer is loaded with the symbols. Compression takes place inside the loop that that iterates until symbols finish. Here the instart is used to keep track of the present bytes being processed in original content and inend is used to keep track of the current byte writing to buffer of the compressed data. During each iteration loop, longest matched cache (LMC) is called to determine the longest match and return the length of it. As the match found, LMC sets the offset to the position of the match in the sliding window and next to the symbol in the look-ahead buffer immediately after the match. In this case, the size of window is 32 KB which determines how far back in the data is searched for matching phrases and limit the length to 258 which allows to find the shorter distances. Hence the smallest distance to reach this length uses only 256 out of 259 for the convenience of array which would make 3 longer. Generally it is good idea to search far back for matchings but it must be balanced against search time through sliding window. Also, balance it against the space penalty by using more bits for offsets. The size chosen for the look-ahead buffer determines the maximum length of phrases that can match. If the data has many long phrases that are duplicated, choosing a buffer size that are too small results in multiple phrase tokens where it might otherwise get just one. The network function htonl is also called to ensure the token in big-endian format. This is the format required to store the compressed data as well as to uncompress it.

## 5. IMPLEMENATION

The proposed algorithm does not need any new inflator that means there is no requirement for new updates at the client-side applications. It is a good approach and viable tool for cutting the cost from heavy traffic websites. The proposed system is implemented in C with the considerations for its easiness and compatibility over the different platforms. Here for the sake of inbuilt libraries, run it on Red Hat Enterprise Linux operating system with kernel 2.6.18-128 (x86_64) on Intel Pentium Dual core CPU E2160 at 1.80 GHz. Also the complete source code has been compiled on GCC version 4.6.3 with a single walled output for the portability, so that it can be directly used anywhere without any pre-configuration. And the benchmarks are selected on the basis of their content features as such Calgary composed of collection of small text with some binary files and Enwik8 which stores 100 million bytes of English Wikipedia large content which are the best for our testing purposes as they have all the necessary content that are always seen in websites while transferring on the HTTP. The only compression libraries are mentioned in it, existing software can be used for their decompression. This provides better functionality with gzip, deflate and compatible with all browsers.

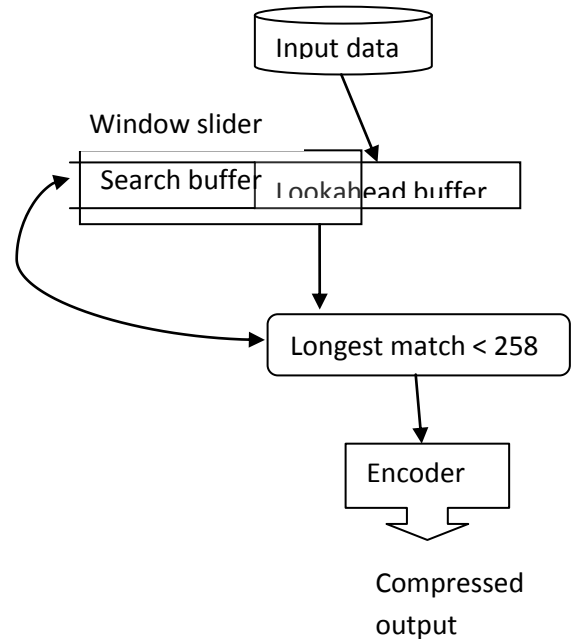

**Fig 2: Deep running of Proposed Algorithm**

## 6. EXPERMINETAL RESULTS

Geflochtener tracks all the backward offsets including even those where no backward references are found and then choose among them that produces the shortest amount of bits. This stores all the lengths and best sequence is found by reverse traversal of the buffer. Some of the corpora has been used for running the compressors: Calgary Corpus [12], Canterbury Corpus [13] and enwik8 [14] which are shown in Table 2.

**Table 2. Comparison of Proposed Geflochtener with existing Compressor**

| Benchmarks | Corpus Size | Gzip-9 | 7Zip | Kzip | Proposed Geflochtener |
|---|---|---|---|---|---|
| Calgary | 3141622 | 1017624 | 980674 | 978993 | 974067 |
| Canterbury | 2818976 | 730732 | 675163 | 674321 | 668456 |
| Enwik8 | 100000000 | 36445248 | 35102976 | 35025767 | 34986660 |

*All sizes in Kilo Bytes.

The compression percentage will be found as per the formulae:

$$C P = \frac{LO - LC}{LO} \times 100$$

where C P is the compression percentage, LO stands for the length of original file and LC stands for length of compressed file. On the contrary, it is also approximated the redundancy rates by 100 – CR%. As from the Table 1, it is depicted that the proposed Geflochtener has removed about 69% redundancy in Calgary which is 1.6% greater than that of gzip-9; similarly for Canterbury it removes 76.22% which is 2.2% better and for English Wikipedia 65% which is 1.5 % greater than that of gzip-9. This yields a remarkable change in compression by removing the redundancies of data about an average of 70%.

Based on the above formulae the performance calculated is shown by graphically in Fig.3 in which the lowest violet line of Geflochtener proves the highest compressibility among the existing compressors.

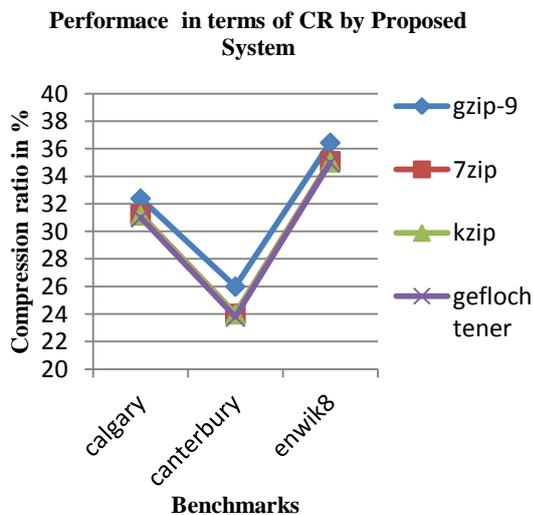

**Fig.3. Comparison of Proposed system with existing compressors**

As discussed, after compressing the data it is need to send over Internet so time recorded while transmitting the compressed Web data over 10 MBps speed connection is depicted in Table 3. whose graph is seen in Fig. 4.

**Table 3. Comparison of Throughput by Proposed Geflochtener over existing Compressor**

| Benchmarks | Gzip-9 | 7Zip | Kzip | Proposed Geflochtener |
|---|---|---|---|---|
| Calgary | 99.4 | 95.8 | 95.6 | 95.2 |
| Canterbury | 71.4 | 65.9 | 65.9 | 65.3 |
| Enwik8 | 3559.1 | 3428 | 3420.5 | 3417.6 |

*All timings in milliseconds

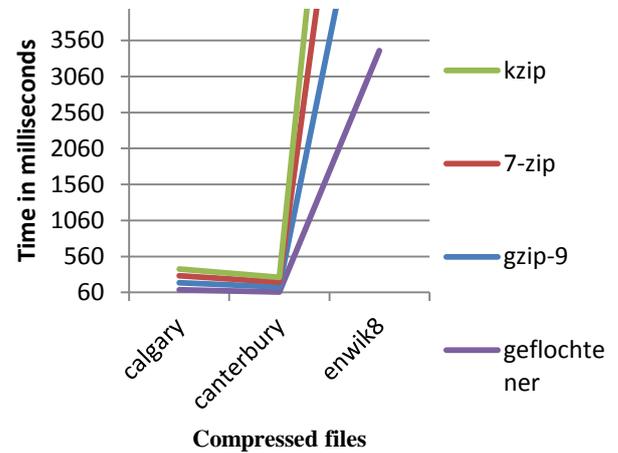

**Fig.4 Comparison of Throughput by Proposed System Compressor**

## 7. CONCLUSION

The proposed system track all the references and takes the best ones based on the greedy implementation with limited iteration to hundred. It does not take the recurring previous lengths while it takes the lengthscore itself for comparison which removes the problem of deciding the length from where we start our next comparison. And the output produced is 4.0-8.52% smaller than that of gzip-9 and save the space on server with the degradation of **514 to 9086** bytes. After performing it on 10 Mbps speed connection, the proposed Geflochtener (with the improvement) transmit Enwik8 content over web in 3 second, Calgary content in 95 milliseconds, and Canterbury content in 65milliseconds which is much significant in accelerating our web traffic. This proves that with the proposed system large data streams of hundreds of **MB can be transmitted within seconds** over the Internet with the help of compression.

The proposed system's strength for binary blobs that change infrequently, if ever, or are downloaded with enough frequency to increase download speed. This also helps in the mobile world where the denser compression results in reduced battery use and less strain on subscriber's data plan. Not only had this it also proved its worth by implementing a stand-in in IDE to create a compact distributable APK files for the users.

As the redundancy brings the vulnerability for cryptanalysis, our Geflochtener compression makes such cryptanalysis harder by reducing the redundancies densely which would be a great benefit while transmitting the encrypted confidential contents after compression over Internet like in Email services.

With all these, it is better phenomenon for a little more compression in not only wired network but also in common wireless spectrum where the mobile data transfers lead to raising the cost to implementation levels. This can further be improved by implementing threading in program to run concurrently [15] [16]. Also it needs to decide when the cleaner should run which will still remains the question of discussion.

## 9. AUTHOR'S PROFILE

**Mr. Hemant Kumar Saini** is a *Red hat Certified Engineer*. He is pursuing M. Tech. in Computer Science & Engineering from Rajasthan Technical University, Kota. He has completed his B. Tech in Information Technology from MLV Government Textile & Engineering College. He is having 2 years of industrial experience and one year of academic experience. He has published articles in CSI and Springer. His research interests are Computer Network, Web Technology and Network Security.

**Mr. Satpal Singh Kushwaha** is an Associate Professor, at MITRC, Alwar (Rajasthan). He has done his M.Tech. from RTU, Kota, B.E. from University of Rajasthan, Jaipur. He has 8 years of teaching and research experience. He has more than 20 papers to his credit, in many international and national journals and conferences. His research interests are Information Security, Network Security and Big Data.

**Dr. Rama Krishna Challa** is an Associate Professor, at NITTTR, Chandigarh. He has done his Ph.D. from IIT Kharagpur, M.Tech. from CUSAT, Cochin and B. Tech from JNTU, Hyderabad. He has 18 years of teaching and research experience. He has more than 75 papers to his credit, in many international and national journals and conferences. His research interests are Wireless Networks, Distributed Computing, Cryptography, and Network Security.